\documentclass{elsart}
\usepackage{amsfonts}
\usepackage{amsmath}

\input{tcilatex}

\begin{document}

\begin{frontmatter}%

\title{Cooperative Parrondo's Games on a Two-dimensional Lattice}%

\author{Zoran Mihailovic and Milan Rajkovic}%

\collab{}%

\address{Institute of Nuclear Sciences Vinca, P.O. Box 522, 11000 Belgrade, Serbia}%

\begin{abstract}
Cooperative Parrondo's games on a regular two dimensional lattice are
analyzed based on the computer simulations and on the discrete-time Markov
chain model with exact transition probabilities. The paradox appears in the
vicinity of the probabilites characterisitic of the "voter model",
suggesting practical applications. As in the one-dimensional case, winning
and the occurrence of the paradox depends on the number of players. 
\end{abstract}%


\end{frontmatter}%

\section{Introduction}

Parrondo's games, inspired by the flashing Brownian ratchet, represent coin
flipping games leading to an apparently paradoxical property that
alternating plays of two losing games can produce a winning outcome \cite%
{Harmer Abbott}. Two types of games, each played by only one player, are
involved denoted as game A and game B. In the former only one biased coin is
used while in the latter two biased coins are used with the player's current
capital determining the state dependent rule. On average, when played
individually each game causes the player to lose. However, when two games
are played in any combination, on average the player always wins. More
specifically, in game A a biased coin with probability of landing head up is 
$p_{A}<1/2$, and assuming that the initial capital is $C=0$, after $n$ plays
the expected value of the capital is $\langle C\rangle=n(2$ $p_{A}-1)<0.$
The game B is played with coins B$_{0}$ and B$_{1}$ with probabilities of
landing head up of $p_{0}$ and $p_{1}$ respectively. The coin B$_{0}$ is
flipped when $C\equiv 0(\func{mod}3)$ and coin B$_{1}$ otherwise. The
quantities of the Brownian ratchets may be made analogous to the variables
figuring in Parrondo's games. For example, the displacement of the particles
corresponds to the capital amount after a certain number of games. The
potential of the Brownian ratchets (usually electrostatic) is analogous to
the games rules, which define the shape of the potential. A comprehensive
review of Parrondo's paradox and references is given in \cite{HarmerAbbott}.

Several variations of these games have been introduced extending the range
of applicability of games possessing the same, apparently paradoxical
property as the original ones. In \cite{Parrondo2}, capital dependence in
game B was replaced by recent history of wins and loses which in turn
inspired the setting in \cite{Torral} where the capital dependence in game B
was replaced by spatial neighbor dependence. The latter games, termed
cooperative, are played by $N$ number of players as a contrast to the
original games which are played by only one player. Each of $N$ players,
arranged in a circle (since periodic boundary conditions are assumed), owns
a capital $C_{i}(t),i=1,...,N$, which evolves by combined playing of games A
and B. Game A is the same as in the original setup, namely it consists of
repeatedly flipping a biased coin A so that the cumulative capital of all
players decreases in time.\ Game B depends on the winning or losing state of
the neighbors on both sides of the player whose turn is to play. It was
shown that the outcome is the same as in the original games, namely that
alternation of games A and B, which may be losing or fair when played
individually, leads to a winning temporal evolution of the total capital of
all players. Recognizing that the spatial dependent games may be played in
an asynchronous manner (each player, chosen randomly or in any other way,
plays when his turn comes) and in a synchronous manner (all players play at
the same time), we have developed a theoretical, discrete-time Markov chain
model of cooperative games with exact rules that allow computation of
transition probabilities for arbitrary number of players \cite{mz1}, \cite%
{mz2}. For these games, termed one-dimensional asynchronous and
one-dimensional synchronous games, rigorous results were obtained for a
small number of players $(N\leq12)$ since analytical expressions increase in
complexity for larger $N$. Exploration of paradoxical properties of
cooperative Parrondo's games naturally extends to the two-dimensional case
which may prove to be of considerable practical importance, surpassing from
that aspect the one-dimensional case. As will be shown in this exposition
two-dimensional games share some properties with the one-dimensional ones,
namely the probabilities for which the paradox occurs depend on the number
of players. Moreover, the paradox occurs for a large number of sets of which
a fairly large quantity exists in the vicinity of probability values
corresponding to the well know voter model\footnote{%
The voter model is a simple mathematical model of opinion formation in which
voters are located at the nodes of a network. Each voter has an opinion (in
the simplest case either 0 or 1), and randomly chosen voter assumes an
opinion of the majority of its neighbors}.

The paper is organized in the following manner: Following a short
presentation of essential rules of the games, we show how a probability
transition matrix may be obtained from the corresponding matrix of the one
dimensional case. Results of computer simulations for the asynchronous and
synchronous cases are presented next and we conclude with suggestions for
possible new directions and applications of these games.

\section{Features of the Games}

\subsection{Rules and mathematical notation}

Each player (or a spin-like particle) may be in one of two states: state $0$
("loser") or state $1$ ("winner"). The state of the ensemble of $M$ players,
in one-dimensional case, may be represented as a binary string $\mathbf{s}%
=(s_{1},...,s_{M}),$ $s_{i}\in(0,1)$ of length $M$, or equivalently, as
state $\mathbf{s}$ in decimal notation. Periodic boundary conditions are
assumed so that $s_{M+1}=s_{1}.$ To each state $\mathbf{s}$ corresponds a
vector equivalent to a basis vector $\left\vert s\right\rangle $ in $P=2^{M}$
dimensional state space 
\begin{equation*}
S_{P}=\{\left\vert s\right\rangle \mid s=0,1,...,P-1\}. 
\end{equation*}
For example, when $M=3$ and $P=8$, state $(011)$ is equivalent to the state $%
\mathbf{3}$, and the corresponding vector is $\left\vert 3\right\rangle
=(00010000)^{T},$ while state $(111)$ is equivalent to state \textbf{7} and
the corresponding vector is $\left\vert 7\right\rangle =(00000001)^{T}.$

It is easy to generalize this representation to a two dimensional $M$ $%
\times N$ array of players located at the nodes of a two-dimensional
lattice, in which case the dimension of the state space becomes $%
P=2^{M\times N}.$ For example in the simplest configuration when there are $4
$ players arranged in a $2\times$ $2$ lattice, the state space is $16$%
-dimensional and the correspondences are illustrated by the following
examples:%
\begin{align*}
\left( 
\begin{array}{cc}
0 & 0 \\ 
0 & 0%
\end{array}
\right) & =(0\text{ }0\text{ }0\text{ }0),\text{ in decimal notation }0;%
\text{ }\longleftrightarrow\text{ }\left\vert \mathbf{0}\right\rangle
=(1000000000000000)^{T} \\
\left( 
\begin{array}{cc}
1 & 0 \\ 
1 & 0%
\end{array}
\right) & =(1\text{ }0\text{ }1\text{ }0),\text{ in decimal notation }10;%
\text{ }\longleftrightarrow\text{ }\left\vert \mathbf{10}\right\rangle
=(0000000000100000)^{T}
\end{align*}
Game A is the same as in the classical setup, while probabilities of winning
in game B depend on the present state of the left and right neighbors
(one-dimensional case) or the four nearest neighbors (two-dimensional case).
For the two-dimensional case five possible configurations of neighboring
players surrounding the player whose turn is to play the game is presented
in Fig. 1. Configuration of neighboring players may be denoted as an ordered
set $(s_{j-1}$ $s_{k-1}$ $s_{j+1}$ $s_{k+1})$, where the indices $k-1,$ $k+1,
$ $j-1$ and $j+1$ denote the state of neighbors to the left, right, up and
below respectively, of a player who is about to play the game. Hence, the
probabilities corresponding to configurations presented in Fig. 1 are:%
\begin{equation*}
\begin{array}{l}
\bullet\text{ }p_{0}^{(B)}\text{ when }(s_{j-1}\text{ }s_{k-1}\text{ }s_{j+1}%
\text{ }s_{k+1})=(0\text{ }0\text{ }0\text{ }0) \\ 
\bullet\text{ }p_{1}^{(B)}\text{ when }(s_{j-1}\text{ }s_{k-1}\text{ }s_{j+1}%
\text{ }s_{k+1})=(0\text{ }0\text{ }0\text{ }1) \\ 
\bullet\text{ }p_{2}^{(B)}\text{ when }(s_{j-1}\text{ }s_{k-1}\text{ }s_{j+1}%
\text{ }s_{k+1})=(0\text{ }0\text{ }1\text{ }1) \\ 
\bullet\text{ }p_{3}^{(B)}\text{ when }(s_{j-1}\text{ }s_{k-1}\text{ }s_{j+1}%
\text{ }s_{k+1})=(0\text{ }1\text{ }1\text{ }1) \\ 
\bullet\text{ }p_{4}^{(B)}\text{ when }(s_{j-1}\text{ }s_{k-1}\text{ }s_{j+1}%
\text{ }s_{k+1})=(1\text{ }1\text{ }1\text{ }1).%
\end{array}
\end{equation*}
\qquad\qquad\qquad It should be remarked that $p_{1}^{(B)}$ applies to any
state with only one winner in the neighborhood, so that, from the aspect of
the player who is about to play, the states $(0$ $0$ $0$ $1),(0$ $0$ $1$ $0),
$ $(0$ $1$ $0$ $0)$ and $(1$ $0$ $0$ $0)$ are equivalent (or four fold
degenerate). States $(0$ $0$ $1$ $1)$ and $(0$ $1$ $1$ $1)$ are six and four
fold degenerate respectively.

\subsection{Evolution of Probabilities}

Winning or losing in any particular game leaves a player in state $1$
("winner") or state $0$ ("loser") respectively. This state remains in effect
until he gets a random chance to play again in the case of asynchronous play
or it changes in each round of synchronous games. Following a play by one of
the players, the state of the ensemble has changed from a state $s(t)$ at
time $t$ to state $s(t+1)$ at time $t+1$. If the probability that an
ensemble in state $s(t)$ (or $\left\vert s(t)\right\rangle )$ is $\pi_{s}(t),
$ then the probability distribution $\pi(t)$ at time $t$ is:%
\begin{equation}
\left\vert \pi(t)\right\rangle =\dsum \limits_{s=0}^{P}\pi_{s}(t)\left\vert
s\right\rangle ,
\end{equation}
while the corresponding probability distribution evolution equation is 
\begin{equation}
\left\vert \pi(t+1)\right\rangle =T\text{ }\left\vert \pi(t)\right\rangle ,
\end{equation}
where $T$ is the probability transition matrix.

Capital $C(t)$ is a function of the ensemble of players which is incremented
by $1$ or decremented by $1$ if one of the players wins or loses
respectively. We introduce the vector of the capital $\left\vert
C\right\rangle $ whose components (their number is equal to the number of
ensemble states) represent the capital generated by each ensemble state.
Each component $C_{s}$ of $\left\vert C\right\rangle $ represents normalized
capital generated by that specific state which is equal to the sum of all
winning and losing individual states in a given ensemble state. Hence, the
state $\mathbf{0}$ generates capital $-1$, while the state $1$ generates
capital $+1$. Explicitly,%
\begin{equation}
C_{s}=\frac{1}{\bar{N}}\dsum \limits_{i=1}^{\bar{N}}(-1)^{s_{i}+1},
\end{equation}
so that the elements of $\left\vert C\right\rangle $ are average values of
the capital generated by each ensemble state separately. For example for $%
\bar
{N}=M\times N=4,$ the vector of the capital is%
\begin{equation}
\left\vert C\right\rangle =1/4\left( -4\text{ }-2\text{ }-2\text{ \ }0\text{ 
}-2\text{ \ }0\text{ \ }0\text{ \ }2\text{\ }-2\text{ \ }0\text{ \ }0\text{
\ }2\text{ \ }0\text{ \ }2\text{ \ }2\text{ \ }4\right) ^{T}.
\end{equation}
The fifth component of $\left\vert C\right\rangle ,$ for example, $-2$
corresponds to the state $\mathbf{4}=(0$ $1$ $0$ $0).$ The ensemble
switching from any state $s(t)$ to the state $s(t+1)=$ $\mathbf{4,}$
generates an average capital $\left\vert C(t+1)\right\rangle =\left\langle
C\mid \mathbf{4}\right\rangle =-2/4=-1/2.$ Furthermore, an ensemble
remaining in state $\mathbf{4}$ throughout its temporal evolution would on
the average generate capital $-1/2$ in each turn of the game. Hence, the
average capital generated by the ensemble is%
\begin{equation}
\left\langle C\right\rangle =\left\langle C\mid\mathbf{\pi}\right\rangle .
\end{equation}
Denoting the probabilities of wining or losing in a certain game by $P_{win}$
and $P_{lose}$ it is easily noticed that%
\begin{align*}
P_{win}+P_{lose} & =1, \\
P_{win}-P_{lose} & =\left\langle C\right\rangle ,
\end{align*}
so that%
\begin{equation}
P_{win}=\frac{1}{2}(1+\left\langle C\right\rangle ).
\end{equation}

\section{Probability Transition Matrix}

\subsection{Asynchronous case}

Probability transition matrix for game A may be easily derived from the one
corresponding to game B, so the later will be derived first. Since at each
moment of time only one player plays and consequently changes the state, the
change may be represented by a Hamming distance between the initial ($i$)
and the final ($f$) state of player $k$:%
\begin{equation*}
d^{H}=\dsum \limits_{s=0}^{\bar{N}}\left\vert i_{k}-f_{k}\right\vert , 
\end{equation*}
where $\bar{N}=M\times N,$ the number of players. Since $d^{H}$ may be
either $0$ or $1$, each case will be considered separately.

\subsubsection{case 1: $d^{H}=0$}

If the ensemble is in state $s$, then the $k$-th player is in state $s_{k}$.
State $s$ of the ensemble also defines the neighborhood $\eta_{k}:(s_{j-1}$ $%
s_{k-1}$ $s_{j+1}$ $s_{k+1})$ of the $k$-th player, which in turn determines
the probability of winning. The ensemble initially in state $i$ may switch
to the state $f=i$ in one of $\bar{N}$ different ways as a result of one of
the players switching from state $i_{k}$ to $f_{k}=i_{k}$, so that the
probability of transition is equal to the sum of probabilities of
independent events%
\begin{equation}
T_{fi}=w(i\rightarrow f)=\frac{1}{N}\dsum \limits_{k=1}^{\bar{N}%
}w(i_{k},f_{k}),   \label{tfi}
\end{equation}
with%
\begin{equation*}
w(i_{k},f_{k})=\left\{ 
\begin{array}{c}
1-p_{\eta_{k}}^{(B)},\text{ \ \ when }f_{k}=0 \\ 
p_{\eta_{k}}^{(B)},\text{ when }f_{k}=1%
\end{array}
\right. 
\end{equation*}

\subsubsection{case 1: $d^{H}=1$}

The $k$-th player switches from state $i_{k}$ to $\ $state $f_{k}$ ($i_{k}$ $%
\neq$ $f_{k})$, with probability%
\begin{equation*}
T_{fi}=w(i\rightarrow f)=\frac{1}{N}w(i_{k},f_{k}). 
\end{equation*}
The size of the probability transition matrix becomes large even for the
case of nine players arranged in a $3\times$ $3$ lattice since the dimension
of the state space is $2^{9}=512,$ imposing a $512$ $\times512$ matrix size.
As in the one-dimensional case this matrix is sparse for the asynchronous
game B while for synchronous game all the elements are non-zero.

\subsubsection{An example}

We illustrate the evaluation of $T_{fi}^{(B)}$ for the simple case of $%
3\times3$ lattice, in the case of transition from state $\mathbf{85}$ to
state $\mathbf{95}$, i.e.%
\begin{equation*}
\left( 
\begin{array}{ccc}
0 & 0 & 1 \\ 
0 & 1 & 0 \\ 
1 & 0 & 1%
\end{array}
\right) =(0\text{ }0\text{ }1\text{ }0\text{ }1\text{ }0\text{ }1\text{ }0%
\text{ }1)\rightarrow\left( 
\begin{array}{ccc}
0 & 0 & 1 \\ 
0 & 1 & 1 \\ 
1 & 1 & 1%
\end{array}
\right) =(0\text{ }0\text{ }1\text{ }0\text{ }1\text{ }1\text{ }1\text{ }1%
\text{ }1). 
\end{equation*}
Evaluation of expression \ref{tfi} for this case yields:%
\begin{gather*}
T^{95\text{ }85}=\frac{1}{9}[prob(0\rightarrow0)+prob(0\rightarrow
0)+prob(1\rightarrow1) \\
+prob(0\rightarrow0)+prob(1\rightarrow1)+prob(0\rightarrow1) \\
+prob(1\rightarrow1)+prob(0\rightarrow1)+prob(1\rightarrow1) \\
=[(1-p_{2})+(1-p_{2})+p_{1}+(1-p_{2})+p_{0}+ \\
p_{3}+p_{1}+p_{3}+p_{2} \\
=\frac{1}{9}(3+p_{0}+2p_{1}-2p_{2}+2p_{3}).
\end{gather*}
Other elements may be derived in a similar manner. The probability
transition matrix for game A is easily obtained by replacing $%
p_{\eta_{k}}^{(B)}$ with $p^{(A)},$ for each $\eta\in\{0,1,2,3\}.$

\subsection{Synchronous case}

Since at each moment of time all players play simultaneously and
consequently change each individual state, the probability of transition is 
\begin{equation*}
T_{fi}=\dprod \limits_{k=1}^{\bar{N}}w(i_{k},f_{k}), 
\end{equation*}
where transition probabilities are 
\begin{equation*}
w(i_{k},f_{k})=\left\{ 
\begin{array}{c}
1-p_{\eta_{k}}^{(B)},\text{ \ \ when }f_{k}=0 \\ 
p_{\eta_{k}}^{(B)},\text{ when }f_{k}=1%
\end{array}
\right. . 
\end{equation*}

\subsubsection{Combination of games A and B}

Combination of games A and B may be played, as in the one-dimensional model,
in two distinctive ways. First, players may simultaneously play game A or
game B in any predetermined or random order, and this case is denoted by
A+B. \ The other possibility is that each player may at each turn of the
game chose randomly whether to play game A or game B, this case being
denoted as A*B. In both cases the paradoxical result that games A+B or A*B
may be winning while each game, A and B, individually losing persists,
however for different sets of probabilities figuring in game B.

\section{Results of simulations}

\subsection{Capital accumulation as a function of time}

In Figs.2 and 3 the averaged capital as a function of time in games A, B and
A+B for the asynchronous and synchronous cases is shown, respectively. In
the synchronous case, Fig. 3, the paradoxical property becomes visible in a
very short amount of time, while the asynchronous case displays paradoxical
property following a short period during which game A fluctuates around a
fair outcome while game B experiences a transient in the positive direction.

\subsection{Probability space defining the paradox}

The results were obtained both by direct numerical simulation and
calculations based on the analytically derived expressions for the
probability transition matrix. As in the one-dimensional setting, the
results show perfect agreement as illustrated in Fig. 4, where the averaged
capital isis shown as a function of probability $p_{1}^{(B)}$ for the case
asynchronously played games$.$ In this diagram a A*B choice for the
alternation of two games is applied. In all simulations the capital
pertaining to the asynchronous games was averaged over 10 000 time steps and
over 1000 runs (ensembles). For synchronous case, since all players play at
each time step, the averaging was performed over 1000 time steps and over
200 runs (ensembles). In the case of calculations based on analytical
expressions for the probability transition matrix, the capital was averaged
as in the case of synchronous games. Simulations of the capital evolution in
the five dimensional probability space show that the paradox occurs for a
very large number of sets and in order to illustrate this, we concentrate on
the appearance of the paradox in the vicinity of the voter model values,
i.e. $p_{0}^{(B)}=0,$ $p_{1}^{(B)}=0.25,$ $p_{2}^{(B)}=0.5,$ $%
p_{3}^{(B)}=0.75$ and $p_{4}^{(B)}=1.$ The obtained probability sets are
displayed in Table 1.

\begin{center}
$\ $%
\begin{table}[tbp] \centering%
$%
\begin{tabular}{ccccc}
\hline
${\small p}_{0}^{(B)}$ & ${\small p}_{1}^{(B)}$ & ${\small p}_{2}^{(B)}$ & $%
{\small p}_{3}^{(B)}$ & ${\small p}_{4}^{(B)}$ \\ \hline\hline
${\small 0}$ & ${\small 0.15}$ & ${\small 0.6}$ & ${\small 0.75}$ & ${\small %
0.95}$ \\ \hline
${\small 0}$ & ${\small 0.15}$ & ${\small 0.6}$ & ${\small 0.8}$ & ${\small %
0.8}$ \\ \hline
${\small 0}$ & ${\small 0.2}$ & ${\small 0.55}$ & ${\small 0.8}$ & ${\small %
0.65}$ \\ \hline
${\small 0}$ & ${\small 0.2}$ & ${\small 0.6}$ & ${\small 0.8}$ & ${\small %
0.85}$ \\ \hline
${\small 0}$ & ${\small 0.2}$ & ${\small 0.6}$ & ${\small 0.7}$ & ${\small %
0.95}$ \\ \hline
${\small 0}$ & ${\small 0.2}$ & ${\small 0.6}$ & ${\small 0.75}$ & ${\small %
0.8}$ \\ \hline
${\small 0}$ & ${\small 0.25}$ & ${\small 0.6}$ & ${\small 0.65}$ & ${\small %
0.95}$ \\ \hline
${\small 0}$ & ${\small 0.25}$ & ${\small 0.6}$ & ${\small 0.7}$ & ${\small %
0.8}$ \\ \hline
${\small 0}$ & ${\small 0.25}$ & ${\small 0.6}$ & ${\small 0.75}$ & ${\small %
0.85}$ \\ \hline
${\small 0.05}$ & ${\small 0.15}$ & ${\small 0.6}$ & ${\small 0.75}$ & $%
{\small 0.9}$ \\ \hline
${\small 0.05}$ & ${\small 0.15}$ & ${\small 0.55}$ & ${\small 0.8}$ & $%
{\small 0.8}$ \\ \hline
${\small 0.1}$ & ${\small 0.2}$ & ${\small 0.6}$ & ${\small 0.7}$ & ${\small %
0.8}$ \\ \hline
\end{tabular}
\ $\caption{Probabilities featured in game B leading to the paradox in the
vicinity of the voter model values}\label{Table 1}%
\end{table}%
\end{center}

As a further illustration, the probability space spanned by ${\small p}%
_{1}^{(B)}$ and ${\small p}_{3}^{(B)}$ and the corresponding paradoxical
range are shown in Fig. 4. Figs. 6, 7, 8, 9 and 10 further illustrate the
paradoxical property of the games through capital evolution as a function of
each of the probabilities determining the state of the player who is about
to play game B. In each of the figures one probability was varied while the
remaining four were kept constant. The region of the paradox is enlarged and
shown in the right diagram.

\subsection{Capital accumulation and lattice size}

In the next two figures we compare the way the capital, as a function of
probability $p_{2}^{(B)}$, depends on the size of the lattice. In Fig. 11,
this dependence is presented for the asynchronous game B. In the cases where
periodic boundary conditions have large influence (small lattice sizes, e.g. 
$3\times3)$ the capital increases linearly and in a noticeable manner. The
same behavior is seen for large lattice sizes (e.g. $100\times100$ and
larger) were periodic boundary conditions do not play an important role.
However for intermediate lattice sizes (from $10\times10$ up to $50\times50$%
) the capital increases substantially only in the (approximate) range $%
0.5\leq$ $p_{2}^{(B)}\leq0.7$. For values on both sides of this range
capital increases very slowly.

For synchronous games this dependence on the lattice size is not distinct,
so that the capital increases, as a function of probability $p_{2}^{(B)},$
in almost the same manner and up to the same value for all lattice sizes
(Fig. 12).

It is interesting that asynchronous game B played at intermediate lattice
sizes leads to the same functional dependence of the capital on the
probability $p_{2}^{(B)}$ as in synchronous games$.$

\subsection{Percolation phase transition}

The dynamics of the games suggests that one should investigate whether there
is a connection between the paradoxical property of the games with the
percolation phase transition. A typical snapshot of the configuration of
"winners" and losers" which arises during the course of the game B is shown
in Fig. 13. In Fig. 14 we superpose two graphs. The first one shows the
capital dependence in games B, A+B and A*B on the probability $p_{2}^{(B)}$,
so that the area of the paradox ( $C^{(B)}<0$ and $C^{(A+B)}>0$ or $%
C^{(A\ast B)}<0)$ is clearly visible. The second graph illustrates the
probability of appearance of the spanning cluster in game B as a function of 
$p_{2}^{(B)}.$ The percolation phase transition occurs at the value of $p_{2%
\mathbf{C}}^{(B)}\simeq0.64$ very close to the range of $p_{2}^{(B)}$ values
for which the paradox occurs (approximately $0.59\leq p_{2}^{(B)}\leq0.62)$
however it is clear that the paradox is not present at $p_{2\mathbf{C}%
}^{(B)}.$ The simulation presented here as a paradigm of the typical
property of the games, was performed for the $50$ $\times$ $50$ lattice size.

\section{Conclusion}

Following a discrete-time Markov chain model of one-dimensional cooperative
Parrondo's games introduced previously, we analyzed the we analyzed the
two-dimensional setting characterized by players arranged at a
two-dimensional regular lattice. It was shown that the alternation of two
losing games, on average, leads to a winning outcome and the number of
probability sets for which this apparently paradoxical property occurs is
very large. In particular, there is a number of probability sets very close
to the values characteristic of the voter model, suggesting possible
applications in the social or economic framework. Moreover, asynchronous
games displaying the paradoxical property are sensitive to the lattice size
(number of players), while this is not the case with the synchronous games.
For medium sized lattices (between $10$ $\times10$ to $50$ $\times50),$ the
capital evolution as a function of one of the probabilities defining game B
behaves in a similar manner for asynchronous and synchronous games.
Percolation phase transition for game B occurs at the probability value very
close to the probability values for which the paradox occurs, however they
seem to be unrelated phenomena. No fronts or patterns were observed at the
paradoxical probability values.

An interesting extension of this work pertains to the games where players
are arranged on a network nodes and where links with the neighbors may be
rewired according to specific rules\cite{MZ}. In this setting an addition of
rewiring probability in game B introduces interesting novel features of the
dynamics along with the existence of the paradox.

\end{document}